\documentclass[amssymb,amsmath,aps,showpacs,twocolumn,floatfix,nofootinbib,showpacs]{revtex4}

\usepackage{graphicx}
\usepackage{color}
\usepackage{soul}
\usepackage{latexsym}

\newcommand{\lsim}{\lesssim}
\newcommand{\gsim}{\gtrsim}
\newcommand{\hin}{h_{inv}}

\def\lsim{\mathrel{\raise.3ex\hbox{$<$\kern-.75em\lower1ex\hbox{$\sim$}}}}
\def\gsim{\mathrel{\raise.3ex\hbox{$>$\kern-.75em\lower1ex\hbox{$\sim$}}}}

\def\beq{\begin{equation}}
\def\eeq{\end{equation}}
\def\beqn{\begin{eqnarray}}
\def\eeqn{\end{eqnarray}}
\def\bea{\begin{eqnarray}}
\def\eea{\end{eqnarray}}
\def\be{\begin{equation}}
\def\ee{\end{equation}}
\newcommand{\fslash}[1]{{#1 \kern -0.7em/ \kern 0.1em}}
\def\ptmiss{\fslash{P}_T}

\begin{document}

\voffset 1.25cm

\title{
Detecting $H\to hh$ in the Mirror Model at the CERN Large Hadron
Collider }

\author{Wen-sheng Li, Peng-fei Yin and Shou-hua Zhu}
\affiliation{Institute of Theoretical Physics, School of Physics,
Peking University, Beijing 100871, China}

\date{\today}

\begin{abstract}

 The Higgs sector may play an important role in detecting the
mirror particles, which can be the candidates of the dark matter and
appear as missing energy in the detectors at the LHC. In this paper
 we worked out the Higgs boson spectrum and the Higgs couplings
for the symmetric vacuum, namely $v_1=v_2=v$, in the mirror model,
and investigated the constraints from electro-weak precision
observable (EWPO). Our study showed that the EWPO has already
constrained the Higgs boson sector severely. We then explored the
Higgs boson phenomenology, and focused on the scenario that the
heavier Higgs boson $H$ can decay into a pair of lighter Higgs boson
$h$. We proposed to study the invisible decay of the Higgs boson via
the pair production of them, in which one Higgs boson decays into
bottom quarks and the other decays invisibly. Our detail simulation
for signals and backgrounds showed that the observation of signal
can reach $5\sigma$ significance for $m_H=260$ GeV and $m_h=115$ GeV
with $10 fb^{-1}$integrated luminosity at the LHC. Moreover the
possible method to further suppress dominant $Zb\bar{b}$ background
was discussed. We also simulated the signals and backgrounds for $H
\rightarrow h h \rightarrow 4b$. Our results showed that it is very
difficult to isolate the signals from huge QCD continuum
backgrounds.

\end{abstract}


\maketitle

\section{Introduction}

Why only the left-handed fermions can feel weak interaction, while
the right-handed fermions can't, is a long-standing fundamental
question in the standard model (SM) of particle physics. Basically
speaking there are two category models to solve this issue. The
first ones are the so-called left-right symmetric models, in which
the SU(2) singlet right-handed fermions in the SM are assumed to
involve the new gauge interaction. Many models of this kind have
been constructed based on the minimal gauge group $SU(2)_L\otimes
SU(2)_R\otimes U(1)_{B-L}$ since the seventies in the last century
\cite{PS:1974,PM:1975}. The second ones are mirror models, in which
the usual SM fermions and/or gauge boson are accompanied by the
mirror partners, similar to the case in the supersymmetric models.
This idea was first proposed by Lee and Yang in 1956 \cite{LY:1956}.
Based on this naive conjecture, many models were constructed. An
excellent extensive review on this subject can be found in Ref.
\cite{O:2006}. One of the advantages of the left-right symmetric
models is that parity can be restored at higher energy. However the
parity restoration scale has been pushed higher and higher, namely
up to several TeVs because of severe constraints from the precise
measurements of the low energy experiments, for example from the
measurement of $K_0-\overline{K}_0$ mixing.

In this paper we will consider the mirror models. Besides providing
a way to understand parity restoration, the mirror particles in the
mirror models can provide {\em natural} candidates for the dark
matter. It is well known that the SM can provide an excellent
description for particle experiments at energies so far probed. But
it can not give any explanation for dark matter. Thus people expect
that new physics beyond the SM can provide a natural candidate for
the dark matter. At present, the most popular candidate for the
(thermal-produced) dark matter is the so-called
weakly-interacting-massive-particle (WIMP). There are some popular
candidates for WIMPs, for example lightest supersymmetric
particle(LSP), axion and lightest Kaluza-Klein particle (LKP) in
universal extra dimension models etc. However the new observations
of dwarf spheroidal galaxies seem indicate that non-cold dark matter
(e.g. sterile neutrino) needs to be considered seriously
\cite{Wyse:2007zw}. Mirror models \cite{F:2007} can also give this
kind of candidate for dark matter.

Besides the cosmological evidences, we need also inputs from
collider experiments, for example the Large Hadron Collider (LHC)
and International Linear Collider(ILC), in order to reach the
decisive conclusions on the properties of dark matter. The basic
assumption here is that the dark matter should couple to usual SM
matter. In mirror models mirror particles can couple to ordinary
particles through three kinds of renormalizable gauge invariant
interactions. The first possibility is through neutrino mixing if
the SM right-handed neutrino, as well as the corresponding mirror
fields, are introduced. If neutrinos have masses, then the mixing
between ordinary and mirror neutrinos is possible. The discussion of
this possibility can be found in Ref. \cite{FV:1995}. The other two
possibilities are 'kinetic mixing' and 'scalar mixing', expressed as
\begin{equation}
L_{mix}=\epsilon
F^{\mu\nu}F^{\prime}_{\mu\nu}+\eta\phi_{1}^{\dagger}\phi_{1}\phi_{2}^{\dagger}\phi_{2}
\label{eq1}
\end{equation}
with $F^{\prime}$ and $\phi_{2}$ the mirror fields of usual U(1)
gauge and Higgs fields respectively in the SM. In Eq. \ref{eq1},
$F^{\mu\nu}F^{\prime}_{\mu\nu}$ will give electric charge to mirror
quarks and leptons. Because QED has been tested very precisely, the
mixing parameter $\epsilon$ is severely constrained to be extremely
small \cite{H:1986,CG:1987,G:2007}, which should play a negligible
role at LHC and ILC. The same conclusion applies also to the
neutrino mixing case. Thus in this paper we will concentrate on the
third possibility: scalar mixing.

As indicated in Eq. \ref{eq1}, mirror Higgs $\phi_{2}$ can couple to
ordinary SM Higgs $\phi_{1}$ through renormalizable gauge invariant
term $\eta\phi_{1}^{\dagger}\phi_{1}\phi_{2}^{\dagger}\phi_{2}$ with
$\eta$ the dimensionless free parameter. After electroweak symmetry
spontaneously breaking, there remain two physical Higgs bosons. Each
Higgs boson is the mixture state of SM and mirror Higgs fields. As a
consequence, each Higgs boson will couple to both usual SM and
mirror particles. So our primary interest in this paper is to
investigate the mirror particle effects on the Higgs sector. The
generic feature of Higgs boson is that Higgs boson can decay
invisibly, i.e. into mirror particles. Such invisible decay can be
searched in all production modes \cite{Z:2006, inv}, and the
promising ones are associated production with Z boson $q \bar q
\rightarrow H Z$, or through gauge boson fusion processes $VV
\rightarrow H$, as well as associated production with top quark
pairs etc.

In this paper we will focus on a new mode to search for the Higgs
boson invisible decay at LHC, i.e. $ gg\rightarrow H\rightarrow hh
\rightarrow b\bar{b}+ mirror\ particles$. Here H and $h$ are the
heavier and lighter Higgs boson in mirror model (see section II).
The mirror particles will appear as the missing energy in the
detectors. The advantage of this process is that light H can be
produced copiously due to the gluon-gluon high luminosity. It is
obvious that $m_H$ is required to be larger than $2 m_h$ in order to
obtain large production rate. As shown below (section III),
$m_H>2m_h$ is allowed in the mirror model after imposing the
constraints from electro-weak precision observable. In the mirror
model, the coupling strength of H-h-h is proportional to
${m_H}^2+2{m_h}^2$ (section II), and the partial decay width of the
mode $H \rightarrow h h$ can be even larger than those that into
gauge bosons. Thus this process can be a promising channel to study
the mirror model. In this paper we will also study the signals and
backgrounds for the process $ gg\rightarrow H\rightarrow hh
\rightarrow b\bar{b}+ b\bar{b}$. Our study shows that it is {\bf a
challenge} to observe this process due to the huge QCD backgrounds.

The paper is organized as following. In section II we briefly
describe the mirror model, and work out the Higgs boson spectrum and
interactions. In section III we investigate the constraints from
electro-weak precision observables utilizing the S, T
parametrization. In section IV we show the main results for the
Higgs boson decays and main production channels. Section V contains
the detail simulation for the signals and backgrounds for the
process $ gg\rightarrow H\rightarrow hh \rightarrow b\bar{b}+
mirror\ particles$ and $ gg\rightarrow H\rightarrow hh \rightarrow
b\bar{b}+ b\bar{b}$. The last section is allocated to discussions
and conclusions.

\section{Mirror Model}

In this section we briefly describe the mirror model (for full
details to see Ref. \cite{FHV:1991}). In the quantum field theory,
the usual parity translation takes $\vec{r}$ to $-\vec{r}$ and $t$
to $t$. In the left-right symmetric models parity is extended to a
new type $Z_2$ discrete symmetry which transforms the left-handed
field to the right-handed one for the {\em same} fermion. Besides
fermion sector, under such $Z_2$ the SM $SU(2)_L$ weak gauge boson
fields transform to the new $SU(2)_R$ gauge bosons and vice versa.
However such discrete symmetry is not solely fixed. Under $Z_2$, the
left-handed sector of fermion field can transform to the
right-handed sector of {\em different} fermion field, namely the
mirror fermion field. Based on this observation, one can extend the
SM by doubling the ordinary fermion, gauge and Higgs fields
\cite{FHV:1991}. Thus the new mirror fermions are natural singlets
of the SM gauge group, and they (nucleus if there exists mirror
$SU(3)_C$) can be the candidates for the dark matter. The minimal
gauge group of the new mirror model is $G_{SM}\otimes G^\prime =
SU(3)\otimes SU(2)\otimes U(1)\otimes SU(3)'\otimes SU(2)'\otimes
U(1)'$. The gauge quantum numbers under $G_{SM}\otimes G^\prime $
for the usual and mirror fermion fields are
\begin{eqnarray}
L^{i}_L\sim(1,2,-1)(1,1,0) &,& (L'_R)^{i}\sim(1,1,0)(1,2,-1)
\nonumber \\
e^{i}_R\sim(1,1,-2)(1,1,0) &,& (e'_L)^{i}\sim
(1,1,0)(1,1,-2) \nonumber \\
Q^{i}_L\sim(3,2,\frac{1}{3})(1,1,0) &,&(q'_R)^{i}\sim
(1,1,0)(3,2,\frac{1}{3})  \nonumber\\
u^{i}_R\sim(3,1,\frac{4}{3})(1,1,0) &,&(u'_L)^{i}\sim
(1,1,0)(3,1,\frac{4}{3}) \nonumber \\
d^{i}_R\sim(3,1,-\frac{2}{3})(1,1,0) &,&(d'_L)^{i}\sim
(1,1,0)(3,1,-\frac{2}{3})\nonumber
\end{eqnarray}
with $i$ the family index.

The $Z_2$ parity symmetry that we define now is
\begin{eqnarray}
\vec{r}\leftrightarrow -\vec{r} , t\leftrightarrow t ,
&G^\mu\leftrightarrow
G'_\mu &, W^\mu\leftrightarrow W'_\mu , B^\mu\leftrightarrow B'_\mu \nonumber \\
L_L\leftrightarrow L'_R , e_R\leftrightarrow e'_L,
&Q_L\leftrightarrow Q'_R& , u_R\leftrightarrow u'_L ,
d_R\leftrightarrow d'_L. \nonumber
\end{eqnarray}

One of the advantages of this model, as we discussed above, is that
there are natural candidates of non-baryonic dark matter in addition
to restore parity \cite{F:2007}. It was shown \cite{F:2007} that
this model does not contradict with the astrophysical and
cosmological observations.

In the following we focus on the Higgs sector of this mirror model.
There are two Higgs fields $\phi_{1}$ and $\phi_{2}$, which are
doublets under $SU(2)$ and $SU(2)^\prime$ respectively. One assumes
that the Higgs potential is invariant under the discrete symmetry
$\phi_{1}\rightarrow\phi_{2}$ to keep the parity in a broader sense.
The Higgs potential is very simply given by
\begin{eqnarray}
V(\phi_{1},\phi_{2})&=&-\mu^{2}\left(\phi_{1}^{\dagger}\phi_{1}+\phi_{2}^{\dagger}\phi_{2}
\right)
+\lambda\left(\phi_{1}^{\dagger}\phi_{1}+\phi_{2}^{\dagger}\phi_{2}
\right)^2 \nonumber\\
&+&\eta\phi_{1}^{\dagger}\phi_{1}\phi_{2}^{\dagger}\phi_{2}.
\end{eqnarray}

After electro-weak symmetry breaking, the Higgs fields can be
written as
\begin{eqnarray}
\phi_i=\left(\begin{array}{cc}
\varphi_{i}\\\frac{1}{\sqrt{2}}(v_{i}+H_{i}+\chi_{i})
\end{array}\right),
\end{eqnarray}
where $\varphi_{i}^{\dagger},\chi_{i}$ are Goldstone bosons, which
will be absorbed by corresponding gauge fields. The vacuum may not
be invariant under $Z_2$ transformation although the Higgs potential
is invariant under this discrete transformation. In fact there are
two ways of spontaneous symmetry breaking, depending on the choice
of the sign of $\eta$ \cite{FLV:2000}. We will discuss these two
cases separately below.

\begin{enumerate}
\item $\eta<0$.

In this case, vacuum is invariant under transformation
$\phi_1\leftrightarrow \phi_2$.
\begin{eqnarray}
v^{2}=v_1^2=v_2^2=\frac{2\mu^{2}}{4\lambda+\eta}.
\end{eqnarray}

We define Higgs boson mass eigenstates as $H$, $h$ (in this paper we
assume $H$ is heavier than $h$),
 \begin{eqnarray}
 H_{1}&=&\frac{1}{\sqrt{2}}(H+h) \\
 H_{2}&=&\frac{1}{\sqrt{2}}(H-h).
 \end{eqnarray}

 The Higgs boson mass can be expressed as
\begin{eqnarray}
 m_{H}^{2}&=&(4\lambda+\eta)v^{2} \\
 m_{h}^{2}&=&-\eta v^{2}.
 \end{eqnarray}

 Obviously
 there must be $\eta<0$, which is coincide with the condition of
 minimizing Higgs potential.

\item $\eta>0$.

For this case, if we require the minimum of Higgs potential is
 stable, then
\begin{eqnarray}
{v_1}^2=\frac{\mu^{2}}{\lambda}, {v_2}^2=0.
 \end{eqnarray}

The Higgs boson masses are

\begin{equation}
{m_h}^2=\frac{\mu^{2}}{2}, m_H^2=\frac{\eta v_1^2}{8}
\end{equation}
It seems that all the mirror particles must be massless. However
mirror particle  can obtain tiny mass through mirror QCD
condensation\cite{FH:1994,FLV:2000}, but we don't discuss this case
further in this paper.

\end{enumerate}

In this paper we only consider the first case, i.e. the vacuum is
invariant under $Z_2$ parity transformation. It should be noted that
the $Z_2$ discrete symmetry of Higgs potential may be broken by
small term. Such term will lead to very different effect
\cite{BTH:2005}.

In order to study the phenomenology of this model, we choose the
Higgs boson masses $m_{H},m_{h}$ as inputs, so
\begin{eqnarray}
\lambda=\frac{m_{H}^{2}+m_{h}^{2}}{4 v^{2}},
\eta=-\frac{m_{h}^{2}}{v^{2}}.
\end{eqnarray}

The Lagrangian of triple interaction of Higgs bosons, which we are
interested in, is given by
\begin{eqnarray}
\mathcal{L}_{int}&=&\omega H h^{2}, \\
\omega &=& \frac{\sqrt{2}}{4}\frac{m_{H}^{2}+2m_{h}^{2}}{v}
\label{triple}
\end{eqnarray}
The other triple and quadruple Higgs interactions can be inferred
from the potential easily which are not shown here.

Thus, if kinematically allowed, $H\rightarrow hh$ decay width can be
expressed as
\begin{eqnarray}
\Gamma(H\rightarrow hh)=\frac{\omega^{2}}{8\pi
 m_{H}^{2}}\sqrt{1-\frac{4m_{h}^{2}}{m_{H}^{2}}}. \label{decaywidth}
\end{eqnarray}

From Eqs. \ref{triple} and \ref{decaywidth}, we can see that the
large decay width is possible for the favorable parameters.

\section{Constraints from Electro-weak Precision Observable}

Current electro-weak precision observable (EWPO) can give
constraints on the physics beyond the SM. A widely used set of
 parameters are  S,T and U \cite{PT:1990}.
 Such parametrization assumes that the new physics contributions
 arise only from propagators of the electro-weak gauge bosons,
 namely the oblique corrections \cite{PT:1990}. The mirror model
 belongs to this case.

 We calculate the Higgs bosons contributions to $S,T$ parameters in
 the mirror model. Here we do not consider $U$ parameter because
 $U$ is very small in general.

We give our results as
 following
\begin{eqnarray}
S=\frac{1}{2} [S_{SM}(m_h)+ S_{SM}(m_H)]- S_{SM}(m_{ref}) \\
T=\frac{1}{2} [T_{SM}(m_h)+ T_{SM}(m_H)]-T_{SM}(m_{ref})
\end{eqnarray}
where \cite{FRW:2001}
\begin{eqnarray}
S_{SM}(m)&=&\frac{1}{\pi} \left[
\frac{3}{8}\frac{m^2}{m_{Z}^2}-\frac{1}{12}\frac{m^4}{m_{Z}^4}+
\frac{m^2}{m_{Z}^2}\log{\frac{m^2}{m_{Z}^2}} \right. \nonumber\\
&\times&\left( \frac{3m_{Z}^2-m^2}{4m_{Z}^2}
+\frac{1}{24}\frac{m^4}{m_{Z}^4}+\frac{3m_{Z}^2}{4(m_{Z}^2-m^2)} \right)\nonumber\\
&+&\left. \left(
1-\frac{1}{3}\frac{m^2}{m_{Z}^2}+\frac{1}{12}\frac{m^4}{m_{Z}^4}\right)
\frac{m}{m_{Z}^2 } f  \right]
\end{eqnarray}
with
\begin{eqnarray}
f &=&
\left\{\begin{array}{cccc}\sqrt{4m_{Z}^2-m^2}\arctan{\sqrt{\frac{4m_Z^2-m^2}{m^2}}}\\if
\;\;\;
m<2m_Z\\
\sqrt{m^2-4m_{Z}^2}\log{\frac{2m_Z}{m+\sqrt{m^2-m_Z^2}}}\\if\;\;\;
m>2m_Z
\end{array}\right.,
\end{eqnarray}
and
\begin{eqnarray}
T_{SM}(m)=\frac{3}{16\pi}\frac{1}{s_{W}^2
c_{W}^2}\left[\frac{m^2}{m_{Z}^2-m^2}\log{\frac{m^2}{m_{Z}^2}} \right. \nonumber\\
\left. -\frac{c_{W}^2 m^2} {c_{W}^2
m_{Z}^2-m^2}\log{\frac{m^2}{c_{W}^2 m_{Z}^2}}\right]
\end{eqnarray}

\begin{figure}[h]
\vspace*{-.03in}
\centerline{\includegraphics[width=3.3in,angle=0]{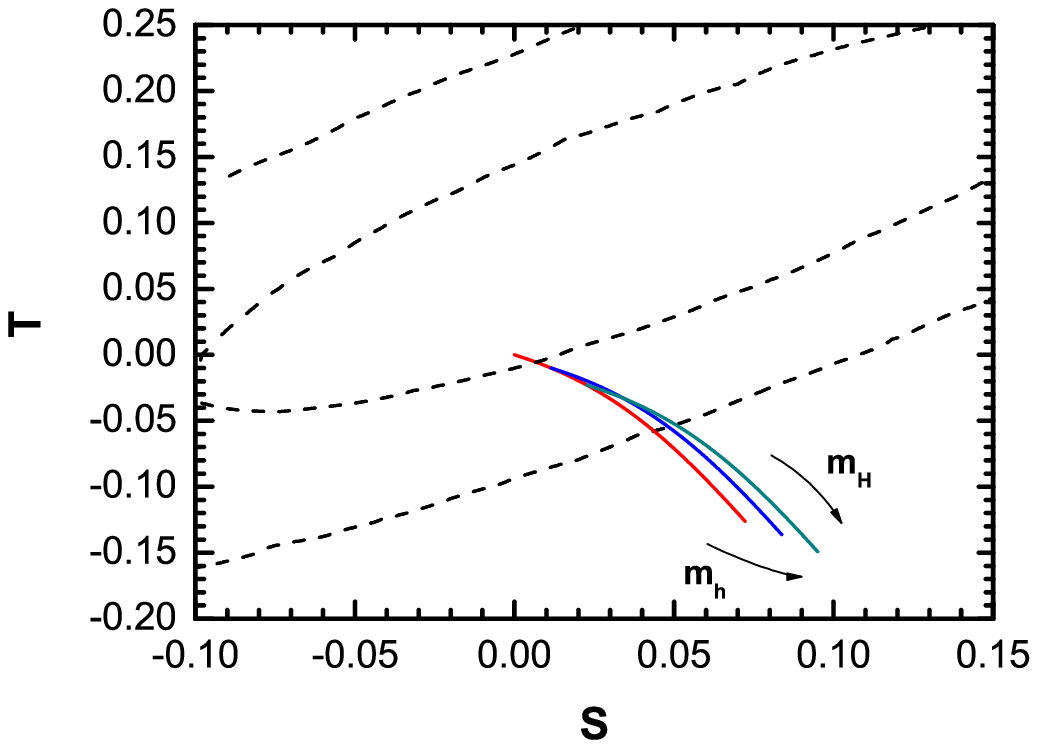}}
\vspace*{-.03in} \caption{The ellipses indicate the regions in the
$S,T$ plane which are allowed by EWPO at $1\sigma$  ($68\%$) and
$2\sigma$ ($95\%$) level respectively \cite{B:2007}. Three curves
represent three different $m_h$ at 115, 150 and 200 GeV, and $m_H$
increases from $100 \sim 1000$ GeV for each curve. \vspace*{-.1in}}
\label{stu}
\end{figure}

The numerical results of S and T are shown in Fig. \ref{stu}. From
the figure we can see that EWPO has already constrained the Higgs
boson masses severely, similar to the SM case. However there is much
freedom than that of the SM. For example, if we fix $m_h=115$GeV,
the 95\% CL upper limit of $m_H$ can approach about 300 GeV. In the
following numerical evaluation we choose $m_H=260$GeV and
$m_h=115$GeV as our benchmark point, except indicated otherwise.

\section{Higgs Boson phenomenology}

In order to discuss the Higgs boson phenomenology, it is convenient
to divide parameters into: (1) $m_{H} < 2 m_{h}$, namely $H$ can't
decay into a pair of $h$, and (2) $m_{H}> 2 m_{h}$.

\subsection{ Higgs Boson Decay}

As mentioned above, each Higgs boson (h and H) is the mixture state
of the ordinary and the mirror Higgs fields. As a consequence, each
Higgs boson can decay into both ordinary and mirror fermions, gauge
bosons if kinematically allowed. In the specific model discussed in
this paper, the branching ratio of the lighter Higgs boson h
decaying into SM particles (fermions and gauge bosons) will be
modified as (taking bottom quark as an example)
\begin{eqnarray}
&&Br(h \rightarrow b \bar b) \nonumber
\\&=& \frac{\Gamma(h \rightarrow b \bar
b)}{\Gamma(h \rightarrow SM)+ \Gamma(h \rightarrow Mirror)} \nonumber \\
&=& \frac{1}{2} Br(h_{SM}\rightarrow b \bar b), \label{eq:20}
\end{eqnarray}
where $Br(h_{SM}\rightarrow b \bar b)$ represents branching ratio of
the SM Higgs boson decaying into bottom quarks. The expressions for
h decaying into mirror particles are the same with those into
ordinary SM particles.

The branching ratios of the heavier Higgs boson H are the same with
those of h in Eq. \ref{eq:20} for the case (1), in which $H
\rightarrow hh$ is kinematically forbidden. Note that we omit here
the tiny three body decay of $H \rightarrow h h^*\rightarrow h f
\bar f$. However for case (2), $H \rightarrow h h$ is allowed. The
formula for the branching ratios of $Br(H\rightarrow hh)$ and
$Br(H\rightarrow b\bar{b})$ are
\begin{eqnarray}
&&Br(H\rightarrow hh) \left[ Br(H\rightarrow b \bar b) \right]
\nonumber
\\&=& \frac{\Gamma(H \rightarrow h h) \left[ \Gamma(H \rightarrow b \bar
b)\right]}{\Gamma(H \rightarrow SM)+ \Gamma(H \rightarrow Mirror)+
\Gamma(H \rightarrow h h)} \nonumber \\
&=&\frac{\Gamma(H \rightarrow h h ) \left[ \Gamma(H \rightarrow b
\bar b)\right]}{2 \Gamma(H \rightarrow SM)+ \Gamma(H\rightarrow h
h)}.
\end{eqnarray}

\begin{figure}[h]
\vspace*{-.03in}
\centerline{\includegraphics[width=3.3in,angle=0]{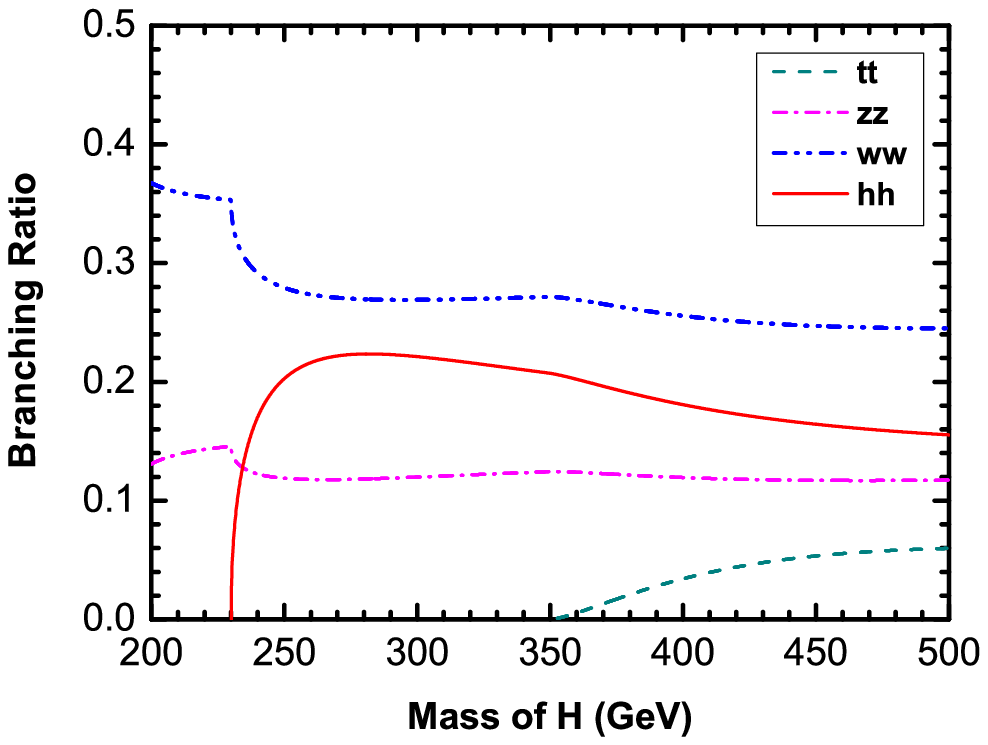}}
\vspace*{-.03in} \caption{Branching ratios of H as a function of
$m_H$, where $m_h=115$GeV. \vspace*{-.1in}} \label{BRofH}
\end{figure}

In Fig. \ref{BRofH}, we show the branching ratios of H to $t\bar t$,
$W^+ W^-$, $ZZ$ and $hh$ as a function of $m_H$ with $m_h=115$ GeV.
From the figure it is obvious the branching ratio of $H \rightarrow
hh$ is larger than that of $Br(H \rightarrow ZZ)$ and sizable once
it is kinematically allowed. Note that $Hhh$ coupling is
proportional to $m_{H}^2+2m_{h}^2$.

If $H$ can decay into a pair of $Z$ boson, it is not difficult to
find such kind Higgs boson via the 'golden' mode with 4 charged
lepton final states. However in the mirror model the branching ratio
of $H\rightarrow ZZ$ can be lower than $20\%$, much higher
luminosity than that of SM is needed to find such kind of the Higgs
boson.

\subsection{ Higgs Boson Production}

In the specific model we discussed in this paper, the production
rate of single Higgs boson h (H) at colliders is only half of that
in the SM. Similar to the SM, the main production channels at the
LHC are $gg \rightarrow h (H)$ or $q q^\prime\rightarrow h(H) V$
with $V=(W, Z)$. For the light Higgs boson, another important
production processes are $gg, qq \rightarrow t\bar t h (H)$. If the
mass difference of two Higgs bosons $h$ and $H$ is larger than that
of detector mass resolution, we may discover two separate mass peaks
out of continuum backgrounds with the much higher luminosity
compared to that in the SM. Note that the branching ratio of h to
the SM particles, as shown in the last subsection, is only half of
that of SM. Similarly more luminosity is needed in order to measure
the properties of the Higgs bosons.

In order to investigate the triple and quadruple couplings among
Higgs bosons, it is necessary to study the pair production of Higgs
bosons. In the mirror model, there are three combinations of the
Higgs boson pair, i.e. $h h$, $h H$  and $H H$. For case (1) it is
hard to investigate such signatures at the LHC due to the low
production rate, similar to that of the SM.

For case (2), $hh$ production can be enhanced due to the s-channel H
resonance. Thus the Higgs boson pair $hh$ production rate can be
much larger than that of case (1). Such enhancement provides one
promising way to study the mirror model. We will focus on this
scenario in the following subsection.

\begin{figure}[h]
\vspace*{-.05in}
\centerline{\includegraphics[width=3.2in,angle=0]{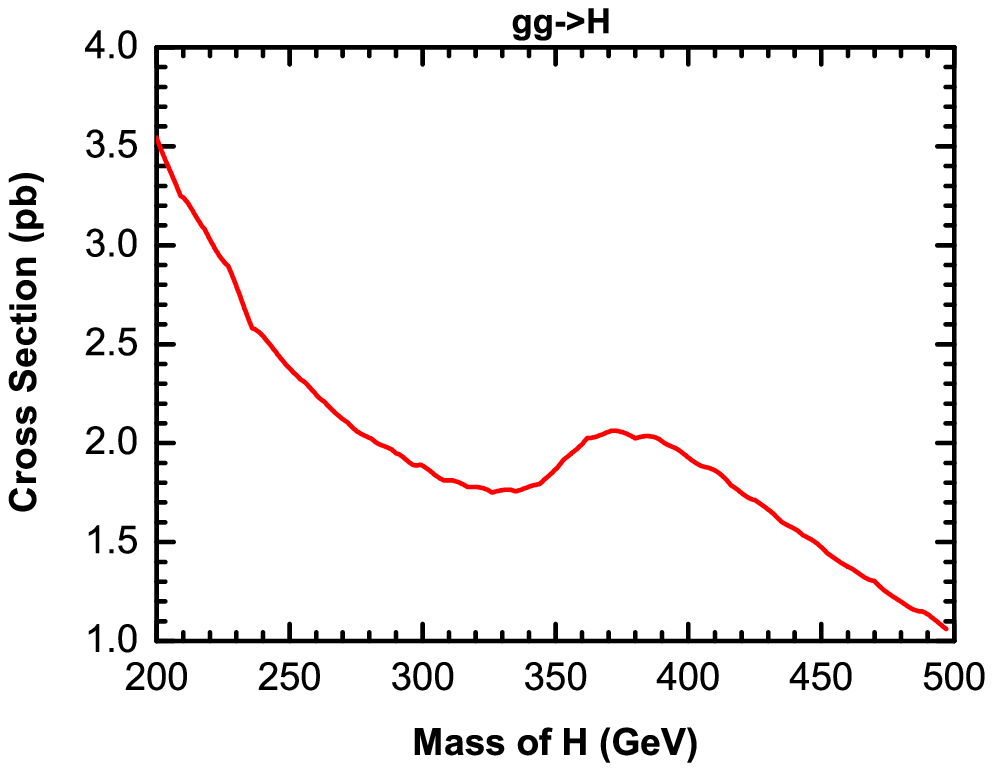}}
\vspace*{-.05in} \caption{Cross sections of $gg\to H$ as a function
of $m_H$ at the LHC. \vspace*{-.1in}} \label{CSofH}
\end{figure}

\begin{figure}[h]
\vspace*{-.05in}
\centerline{\includegraphics[width=3.2in,angle=0]{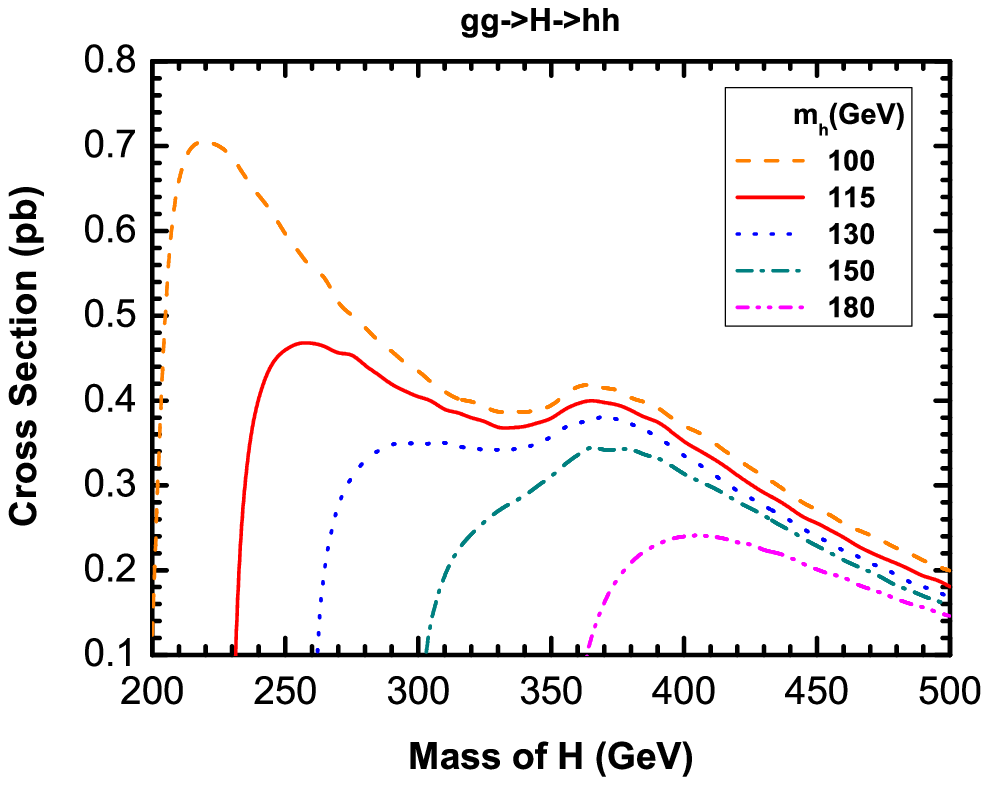}}
\vspace*{-.05in} \caption{Cross sections [in pb] of $gg\to H \to hh$
as a function of $m_H$ at the LHC,where $m_h=100,115,130,150 $ and
$180$ GeV from top to bottom . \vspace*{-.1in}} \label{5CSofH}
\end{figure}

In Fig. \ref{CSofH} and Fig. \ref{5CSofH} we show the cross sections
for the single H and hh pair production as a function of $m_H$. From
the figures it is obvious that the hh pair production can easily
reach several hundred of femtobarns.

In literature there are extensive investigations on processes $H
\rightarrow hh \rightarrow $ light quarks and/or leptons and/or
$\gamma\gamma$ \cite{Zhu:2006zv, BCW:2007, Haa, SWW:2006, CSY:2007},
where $h$ represents light scalar or pseudoscalar. If  $h$ is heavy
enough, the channel of $H\to hh \to 4W$ would open \cite{BPR:2002}.
However such scenario is not favored by the precision data, as shown
in section III. In the mirror model, in order to maintain the signal
rate, we require that one lighter Higgs boson $h$ decay into
$b\Bar{b}$, and the other lighter Higgs boson has many decay modes.
In this paper we focus on two modes: (1) h decays into mirror
particles which has the largest branching ratio of 50\%; and (2) h
decays into $b\bar b$ which has the second largest branching ratio,
larger than that of $l^+l^-$ or $\gamma\gamma$ etc.

It should be mentioned the other possible channel, namely $q
q^\prime \rightarrow V + H\rightarrow \ell + 4b$, which can also be
utilized to study $H \rightarrow hh$ \cite{CSY:2007}. As shown in
Ref. \cite{CSY:2007}, for light Higgs boson with the mass around 100
GeV, this process may provide a clean signature out of the
backgrounds. But for $m_{H}=260$ GeV, there are large background
from $gg \rightarrow tt$ with $t\bar t \rightarrow \ell+ 2b+2j$, and
from light quark jets with two of them misidentifying as b jets.

\section{Detail Simulations}

In this section, we will simulate the signals and backgrounds for
the process $gg \rightarrow H\rightarrow hh$ choosing $m_{H} = 260
$GeV and $m_{h} =115 $GeV as the typical parameters. We utilize
Madgraph/MadEvent 4.1.27 \cite{MGME:2007} and Pythia 6.4.11
\cite{pythia:2006} to simulate the backgrounds and signals at the
LHC with the CTEQ5L PDF set. We ignore initial and final QCD and QED
radiation corrections. For the sake of simplicity, we assume the
b-tagging efficiency of $50\%$ and mis-tagging efficiencies for c, g
and light quarks of $10\%$ , $1\%$ and $1\%$, respectively
\cite{SWW:2006}.

\subsection{$gg \rightarrow H\rightarrow hh \rightarrow b\overline{b}+Mirror\  Particles $}

In the mirror model, the largest branching ratio of $h$ is the decay
into mirror particles. Because the mirror particles appear as the
missing energy at colliders, such decaying $h$ can't be directly
reconstructed via its decay products, i.e. h decays invisibly. In
order to investigate $H \rightarrow hh$, we require that the other h
must decay into bottom quarks, which is the largest {\em visible}
decay mode for our choice of $m_h$.

As pointed out in the Introduction, the invisible Higgs boson has
been extensively investigated in literature \cite{MW:1999, Z:2006,
inv}. In this section, we investigate the observability of invisible
Higgs boson at the LHC, via the process
\begin{equation}
 p\ p \to  g \, g \to H \to h(\to b^+b^-) + \hin,
 \label{pptozh2}
\end{equation}
where one $h$ decays into bottom quarks and the other decays
invisibly. As it was known \cite{MW:1999}, one of the disadvantages
of this $b\bar{b}+ \ptmiss$ signal is that $b\bar{b}$ final states
are not easy to be reconstructed completely.

The most important irreducible background arises from $Zb\bar{b}$
production
\begin{equation}
 p \, p \to Z(\to \nu \bar{\nu})b\bar{b}, \label{pptozh3}
\end{equation}
where $Z$ decays into neutrinos. Moreover QCD multi-jet production,
such as $p \, p \to Z(\to \nu \bar{\nu})jj$, are also the sources of
the large backgrounds. In this paper we require two b-tagged jets in
order to suppress these backgrounds. Other backgrounds can arise
from $ZZ$, $WZ$, $Wb\bar{b}$, single top and $t\bar{t}$ production
\cite{MW:1999}.

We utilize MadEvent \cite{MGME:2007} to simulate all backgrounds,
and apply the basic kinematical cuts as following
\begin{eqnarray}
& P_T(j_1),P_T(j_2) > 20GeV , 15GeV & \label{c1}\\
& |\eta_j| < 2 & \label{c2} \\
& \triangle R (jj) > 0.4 & \label{c3}\\
& m_{jj} > 10 GeV, & \label{c4}
\end{eqnarray}
where $P_T$ denotes the transverse momentum of jet, $\eta$ denotes
the pseudo-rapidity, and $\Delta R
=\sqrt{(\Delta\eta)^2+(\Delta\phi)^2}$ with $\phi$ the azimuthal
angle.

In the following we discuss how to suppress the backgrounds from
$Wb\bar{b}$, $WZ$, single top with $t\to b W$ and
$t\bar{t}\rightarrow b W \bar b W$. For these backgrounds, the final
state charge leptons or jets from $W$ escape from the detection. We
suppress these contributions by vetoing events from $W$ decay with
follow cuts
\begin{eqnarray}
& P_T(j)>15 GeV , |\eta(j)|<2.0 & \label{bc5} \\
& P_T(l^\pm)>10 GeV , |\eta(l^\pm)|<2.5 & \label{bc6}
\end{eqnarray}

The numerical results after imposing cuts Eqs. \ref{c1}-\ref{bc6}
are shown in Tab. \ref{table1}. It is obvious that the dominant
backgrounds come from the $Zb\bar b$ production.

\begin{table}[htb]
\begin{tabular}{l|cccccc}
\hline \hline
Channel &\,\,\, $Zb\bar{b}$  &\,\,\, $Zb\bar{c}$  &\,\,\, $Zbj$  &\,\,\,  $Zc\bar{c}$  &\,\,\, $Zcj $  &\,\,\, $Zjj$  \\
\hline
\,\,\,$\sigma(pb)$ &\,\,\,3.250   &\,\,\,0.011   &\,\,\,0.107   & \,\,\,0.001   &\,\,\,0.027   &\,\,\,0.063  \\
\hline \hline
Channel &\,\,\,  $ZZ$  &\,\,\, $W^-b\bar{b}$  &\,\,\,  $W^-Z$  &\,\,\,  $t\bar{b}$  &\,\,\, $t\bar{t}$  \\
\hline
\,\,\,$\sigma(pb)$ & \,\,\,0.072  &\,\,\,0.417  &\,\,\,0.032  &\,\,\,0.017  & \,\,\,0.346 \\
\hline \hline
\end{tabular}
\caption{The cross sections (in pb) of backgrounds for
$b\bar{b}+\ptmiss$ after basic kinematical cuts Eqs.
\ref{c1}-\ref{bc6} and tagging efficiencies where
$j=u,\bar{u},d,\bar{d},s,\bar{s},g$.}\label{table1}
\end{table}

\begin{figure}[h]
\vspace*{-.03in}
\centerline{\includegraphics[width=3.3in,angle=0]{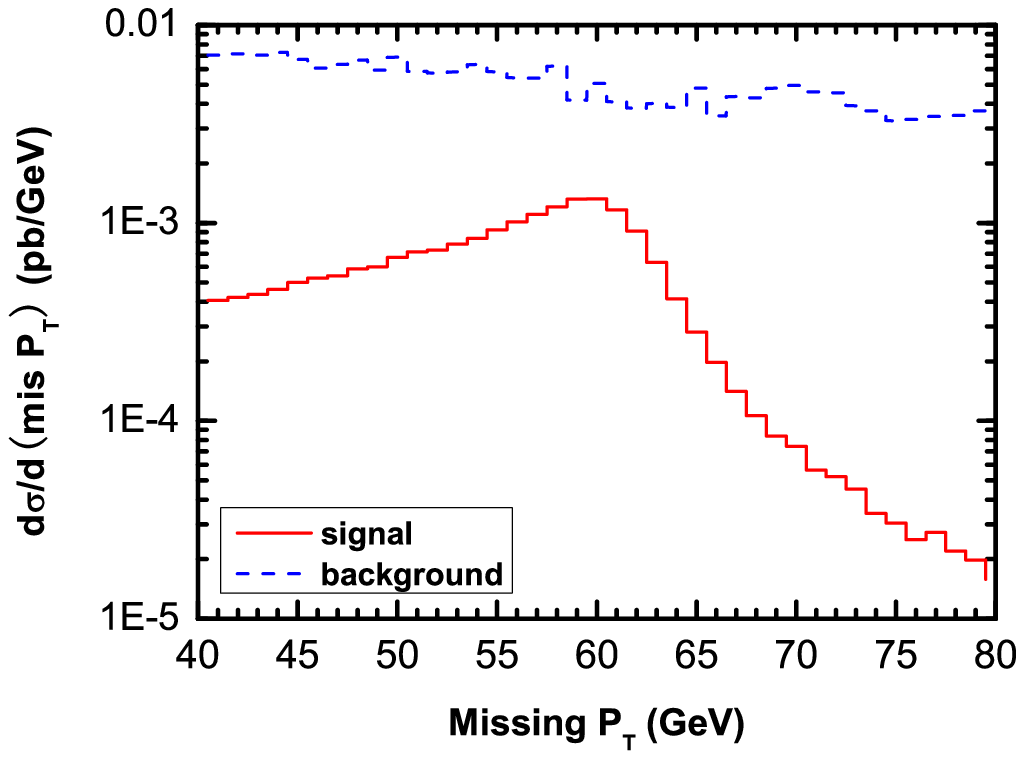}}
\vspace*{-.03in} \caption{The distributions of the signals and
backgrounds for $b\bar{b}+\ptmiss$ as a function of $\ptmiss$ after
applying cuts Eqs \ref{c1}-\ref{bc6} and tagging efficiencies.
\vspace*{-.1in}} \label{mpt1cs1bc}
\end{figure}

The signals and backgrounds for $b\bar{b}+\ptmiss$ as a function of
$\ptmiss$ are shown in Fig. \ref{mpt1cs1bc}, where we reconstruct
$\ptmiss$ from $b\bar{b}$. We can see that the signals peak within
the $40$GeV$<\ptmiss <80$GeV window, while the backgrounds are flat.
In order to suppress the backgrounds, we impose the further cuts as
following
\begin{eqnarray}
& |m_{jj}-m_h|<15GeV &\label{c5} \\
& 40 GeV <\ptmiss < 80 GeV. \label{mr}&
\end{eqnarray}
It is natural to impose the cut of Eq. \ref{c5} because we can
extract the approximate mass information of Higgs boson via other
process, for example $gg \rightarrow h \rightarrow \gamma \gamma$.

\begin{figure}[h]
\vspace*{-.03in}
\centerline{\includegraphics[width=3.3in,angle=0]{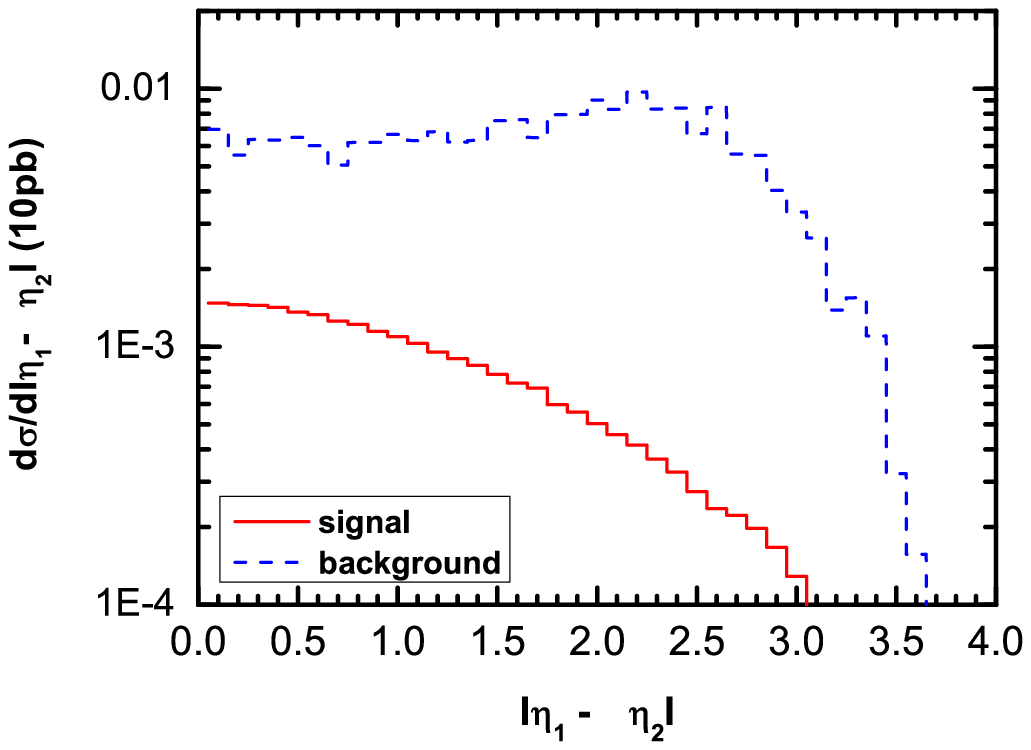}}
\vspace*{-.03in} \caption{The distributions of the signals and
backgrounds for $b\bar{b}+\ptmiss$ as a function of $
|\eta_{j_1}-\eta_{j_2}|$ after imposing cuts Eqs. \ref{c1}-\ref{mr}
and tagging efficiencies . \vspace*{-.1in}} \label{det1cs1bc}
\end{figure}

We show the signals and backgrounds as the function of $
|\eta_{j_1}-\eta_{j_2}|$ after imposing all above cuts Eqs.
\ref{c1}-\ref{mr} in Fig. \ref{det1cs1bc}. The figure shows that
signals decrease quicker than that of backgrounds for
$|\eta_{j_1}-\eta_{j_2}|<2$, which provides a possible way to
suppress backgrounds further. We require
\begin{equation}
|\eta_{j_1}-\eta_{j_2}| < 1.5 \label{etacut},
\end{equation}
and this cut would improve significance of the signals by a factor
of $1.2$, as shown explicitly in Tab. \ref{table2}.

From the Fig. \ref{det1cs1bc}, we can see that the backgrounds are
still much larger than those of the signals. The dominant
backgrounds come from the irreducible $Zb\bar{b}$ production. We
would like to explore another additional potentially useful methods
to reduce the this huge background, similar to the case in Ref.
\cite{Z:2006}.

In order to suppress the largest $Zb\bar{b}$ background, we can
utilize the precise measurement of $Z(\to \mu^+\mu^-) b \bar b$,
similar to that in Ref. \cite{Z:2006}. The reason is that the
backgrounds  $Z (\to \nu \bar \nu)b\bar{b}$ have similar kinematics
and distributions to those of $Z(\to \mu^+\mu^-) b\bar{b}$
production. We can obtain improved (reduced) background $Zb\bar{b}$
as following
\begin{equation}
\sigma_{bkg}^{Zb\bar{b},imp}=\sigma_{bkg}^{Zb\bar{b}}-R\times\sigma_{b\bar{b}\mu^+\mu^-}.\label{q1}
\end{equation}
In Eq. \ref{q1}, $\sigma_{b\bar{b}\mu^+\mu^-}$ is the cross section
for $Z(\to \mu^+\mu^-)b\bar{b}$ production which adopts the same
kinematical cuts for $Z(\to \nu \bar \nu)b\bar{b}$. $R$ is a ratio
which is  defined as
\begin{equation}
R=\frac{\sum_i Br(Z\to\nu_i\bar{\nu_i})}{Br(Z\to\mu^+\mu^-)},
\end{equation}
and in our case $R=5.94$ \cite{Z:2006}. Note that
$\sigma_{bkg}^{Zb\bar{b}, imp} \approx 0$ if we can measure all
final states $b\bar{b}\mu^+\mu^-$ in any kinematical region.

However we are aware of the difficulties of this method. For
example, in order to combine two measurements, at least we must deal
with two different systematics, and fully understand the detectors
etc. Detail study on this method is obviously beyond the scope of
this paper, and we show here only the rough estimation.

\begin{table}[htb]
\begin{tabular}{l|ccccc}
\hline \hline
\,\,\,\,\,\,\,\,\,\,\,\,\,\,\,\,\,\,\,Cuts &$s(fb)$ &$b(fb)$  &$S/B$  &$S/\sqrt{B}_1$ &$S/\sqrt{B}_2$   \\
\hline
\,\,\,\,\,\,\,\,\,\,\,\,\,basic cuts &26.6 &4948 & 0.0054  &1.19  &2.07    \\
\hline
$|m_{jj}-m_h|<30$GeV &26.6 & 1133 & 0.023  &2.50  & 4.32   \\
\hline
$|m_{jj}-m_h|<15$GeV &26.6 & 492 &  0.054  &3.79  & 6.56   \\
\hline
\,\,\,$20<\ptmiss<120$GeV &25.0 &401 & 0.062  &3.94 &6.83    \\
\hline
\,\,\,\,$40<\ptmiss<80$GeV &19.4 & 202 & 0.096  &4.33 &7.49   \\
\hline
\,\,\,\,\,\,$|\eta_{j_1}-\eta_{j_2}|<1.5$ &15.2 &95 & 0.16  &4.93  &8.54  \\
\hline \hline
\,\,\,\,\,\,improved backg &15.2 & 18 & 0.83  &11.4  &19.8    \\
\hline \hline
\end{tabular}
\caption{The effects on the cross sections of signal (s) and
background (b), ratio of signal and background events $S/B$,
$S/\sqrt{B}_1$ and $S/\sqrt{B}_2$, by imposing cuts of
Eqs.\ref{c1}-\ref{q1} step by step, are summarized. Here $m_H=260$
GeV and $m_h=115$ GeV. The significance $S/\sqrt{B}_1$ is for the
luminosity of $10fb^{-1}$ and $S/\sqrt{B}_2$ is for the luminosity
of $30fb^{-1}$. All the numbers shown here are after tagging
efficiencies. \vspace*{-.1in}}\label{table2}
\end{table}

All numerical results after imposing cuts step by step for Eqs.
\ref{c1}-\ref{q1} are summarized in Tab. \ref{table2}. From Tab.
\ref{table2}, we can see that at the LHC, the channel $H\to hh \to
b\bar{b}+\ptmiss$ can provide about $5\sigma (11\sigma)$
significance observation with only $10fb^{-1}$ after applying the
cuts Eqs. \ref{c1}-\ref{c5} and suitable $\ptmiss$ window, without
(with) further suppressing the $Zb\bar{b}$ backgrounds using Eq.
\ref{q1}. In other words, it is not difficult to detect $H
\rightarrow hh \rightarrow b\overline{b}+ mirror\ particles$ at the
LHC. To see this explicitly, we give the luminosity (Tab.
\ref{table3}) which is required to achieve $5\sigma$ observation for
different $m_H$ and $m_h$.

\begin{table}[htb]
\begin{tabular}{l|c|c|c}
\hline \hline
& $m_h=100$GeV & $m_h=115$GeV & $m_h=130$GeV  \\
\hline
 $m_H=250$ GeV &\,\,8.2(40,80) &\,\,8.3(10,60) &\,\,$--$ \\
\hline
$m_H=300$ GeV &\,\,9.0(80,130) &\,\,9.6(60,110) &\,\,17.5(40,80)  \\
\hline
$m_H=350$ GeV &\,\,5.5(100,150) &\,\,6.6(90,140)&\,\,11.6(80,120)   \\
\hline \hline
\end{tabular}
\caption{The integrated luminosity [in $fb^{-1}$], which is required
to observe $H\to hh \to b\bar{b}+\ptmiss$ with $5\sigma$
significance at the LHC, for several sets of $m_H$ and $m_h$. The
numbers in bracket are mass window of $\ptmiss$. Note the Eq.
\ref{q1} is not applied. \vspace*{-.1in} }\label{table3}
\end{table}

\subsection{$gg \rightarrow H\rightarrow hh \rightarrow 4b $}

In Ref. \cite{SWW:2006}, the authors studied the general 4b signal
in extended SUSY models (with $p_T^j> 15$ GeV ) at Tevatron. They
concluded that for the SM case $gg\rightarrow h$ has not enough
rate, an order of magnitude smaller, for the light SM Higgs boson
with mass less than 150 GeV. However if the signal is sufficiently
enhanced, $4b$ signal may be detectable. As we have shown, the $hh$
production rate can be greatly enhanced once $H \rightarrow hh$ is
allowed. In this section we use the same strategy to investigate
$4b$ signals and backgrounds as those in Ref. \cite{SWW:2006}.

Most of the backgrounds for $gg\to H\to hh\to 4b$ are due to the
contributions from large QCD multijet production. We adopt basic
kinematical cuts as Eqs.(\ref{c1})$\sim$(\ref{c4}). Moreover we
require at least tagging three b-jets to reduce the huge
backgrounds. The backgrounds drop from $0.106 mb$ to $3.21 nb$. In
order to suppress the backgrounds further, we require
$|m_H-m_{4j}|<$20 GeV and $|m_h-m_{jj}|<$15 GeV.

\begin{figure}[h]
\vspace*{-.03in}
\centerline{\includegraphics[width=3.3in,angle=0]{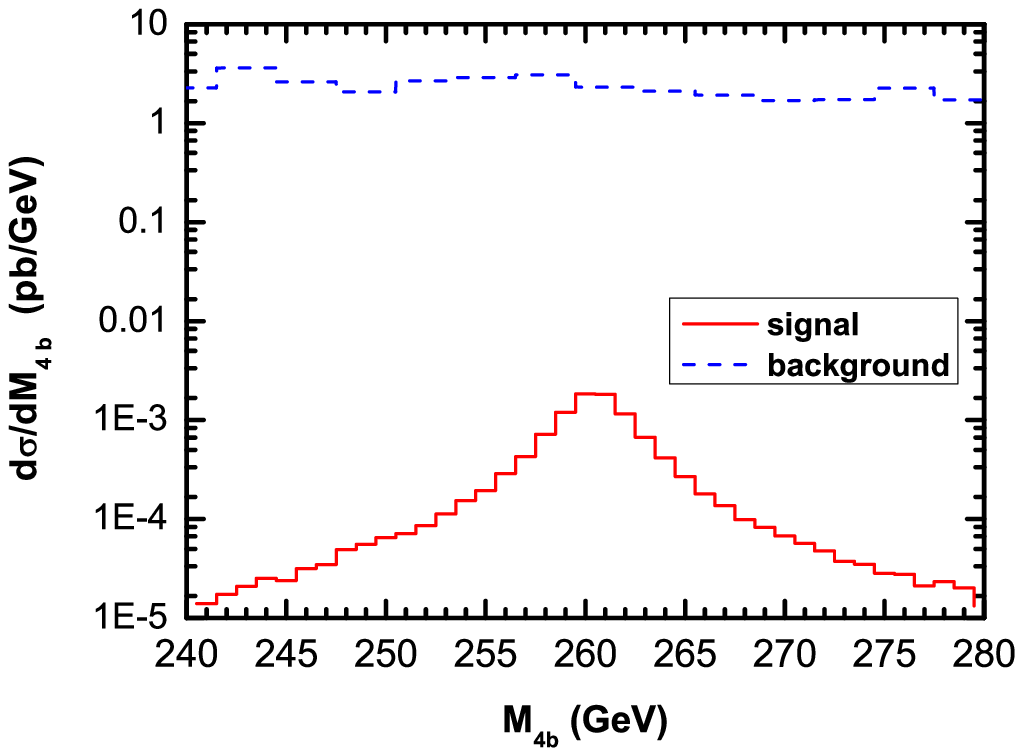}}
\vspace*{-.03in} \caption{The distributions of the signals and
backgrounds for $b\bar{b}b\bar{b}$ as a function of invariant mass
for $4b$ after applying cuts (see text) and tagging efficiencies.
\vspace*{-.1in}} \label{bbbb}
\end{figure}

In Fig. \ref{bbbb} we show the signals and backgrounds as a function
of invariant mass of four jets. It is obvious that the backgrounds
are much larger, around three orders of the magnitude, than those of
the signals. The main reason is that the gluon-gluon luminosity
drops very quickly with the increment of the $m_H$.

\section{Discussions and conclusions}

 The Higgs sector may play an important role in detecting the
mirror particles, which can be the candidates of the dark matter and
appear as missing energy in the detectors at the LHC. In this paper
 we worked out the Higgs boson spectrum and the Higgs couplings
for the symmetric vacuum, namely $v_1=v_2=v$, in the mirror model
\cite{FHV:1991}, and investigated the constraints from electro-weak
precision observable (EWPO). Our study showed that the EWPO has
already constrained the Higgs boson sector severely. We then
investigated the Higgs boson phenomenology, and focused on the
scenario that the heavier Higgs boson $H$ can decay into a pair of
lighter Higgs boson $h$.

As the generic feature in the mirror model, Higgs boson can decay
invisibly. The invisible Higgs boson can be detected via the
processes \cite{Z:2006, inv} of $q \bar q \rightarrow ZH$, weak
gauge boson fusion $VV \rightarrow H$ or $gg \rightarrow t\bar t H$
etc. In this paper we proposed to study the invisible decay of the
Higgs boson via pair production of them, in which one Higgs boson
decays into bottom quarks and the other decays invisibly. Our detail
simulation for signals and backgrounds showed that the observation
of signal can reach $5\sigma$ significance for $m_H=260$ GeV and
$m_h=115$ GeV with $10 fb^{-1}$integrated luminosity at the LHC. It
should be emphasized that our conclusion also applies to other
models in which the Higgs boson pair production cross section is
O(100) femtobarns and the branching ratio of invisible decay is
sizable. Moreover the possible method to further suppress dominant
$Zb\bar{b}$ background was discussed.

We also simulated the signals and backgrounds for $H \rightarrow h h
\rightarrow 4b$ at the LHC. Our results showed that it is very
difficult to isolate the signals from huge QCD continuum
backgrounds.

In the mirror model, the heavier Higgs boson $H$ may not decay into
a pair of lighter Higgs boson $h$. For such case, it is hard to
study the triple Higgs couplings of hhH. In fact even $H \rightarrow
hh$ is allowed, the other triple Higgs couplings such as hhh and HHH
are very difficult to measure at the LHC. Therefore  ILC is the
necessary next facility of investigating Higgs triple, even
quadruple couplings.

\section{Acknowledgements}

SHZ thanks Prof. K.T. Chao and Prof. Y.P. Kuang for drawing the
attention to the issue of parity restoration.
 This work was supported in part by the Natural Sciences Foundation of
China (No. 90403004, 10775001
 and 10635030), the trans-century fund and the
key grant project (No. 305001) of Chinese Ministry of Education.

\end{document}